\documentclass[lettersize,journal]{IEEEtran}
\usepackage{amsmath,amsfonts}
\usepackage{algorithmic}
\usepackage{algorithm}
\usepackage{array}
\usepackage[caption=false,font=normalsize,labelfont=sf,textfont=sf]{subfig}
\usepackage{textcomp}
\usepackage{stfloats}
\usepackage{url}
\usepackage{verbatim}
\usepackage{graphicx}
\usepackage{cite}

\usepackage{siunitx}
\begin{document}

\title{Validating Computational Markers of Depressive Behavior: Cross-Linguistic Speech-Based Depression Detection with Neurophysiological Validation}

\author{Fuxiang Tao, Dongwei Li*, Shuning Tang*, Xuri Ge, Wei Ma, Anna Esposito, \\
Alessandro Vinciarelli,~\IEEEmembership{Member,~IEEE}
        % <-this % stops a space
\thanks{The research leading to these results has received funding from the project ANDROIDS funded by the program V:ALERE 2019 Università della Campania “Luigi Vanvitelli”, D.R. 906 del 4/10/2019, prot. n. 157264,17/10/2019. The work of Alessandro Vinciarelli was supported by UKRI and EPSRC through grants EP/S02266X/1 and EP/N035305/1, respectively. The work of Dongwei Li was supported by the National Natural Science Foundation of China (32400863), and the Open Research Fund of the State Key Laboratory of Brain-Machine Intelligence, Zhejiang University (BMI2400006).}% <-this % stops a space
\thanks{Fuxiang Tao, Dongwei Li and Shuning Tang are with Department of Psychology, Faculty of Arts and Sciences, Beijing Normal University, Zhuhai, 519087, China; Beijing Key Laboratory of Applied Experimental Psychology, National Demonstration Center for Experimental Psychology Education, Faculty of Psychology, Beijing Normal University, Beijing, 100875, China.}
\thanks{Xuri Ge is now with the School of Artificial Intelligence, Shandong University, 250100, China.}
\thanks{Wei Ma and Alessandro Vinciarelli are with the School of Computing Science, University of Glasgow, G12 8RZ, UK. Xuri Ge is now with the School of Artificial Intelligence, Shandong University, 250100, China.}
\thanks{Anna Esposito is with the Dipartimento di Psicologia, Università della Campania “Luigi Vanvitelli”, Caserta, 81100, Italy.}
\thanks{* Corresponding authors: Dongwei Li and Shuning Tang.}}

\markboth{Journal of \LaTeX\ Class Files,~Vol.~14, No.~8, August~2021}%
{Shell \MakeLowercase{\textit{et al.}}: A Sample Article Using IEEEtran.cls for IEEE Journals}


\maketitle

\begin{abstract}
Speech-based depression detection has shown promise as an objective diagnostic tool, yet the cross-linguistic robustness of acoustic markers and their neurobiological underpinnings remain underexplored. 
This study extends Cross-Data Multilevel Attention (CDMA) framework, initially validated on Italian, to investigate these dimensions using a Chinese Mandarin dataset with Electroencephalography (EEG) recordings. We systematically fuse read speech with spontaneous speech across different emotional valences (positive, neutral, negative) to investigate whether emotional arousal is a more critical factor than valence polarity in enhancing detection performance in speech. Additionally, we establish the first neurophysiological validation for a speech-based depression model by correlating its predictions with neural oscillatory patterns during emotional face processing. 
Our results demonstrate strong cross-linguistic generalizability of the CDMA framework, achieving state-of-the-art performance (F1-score up to 89.6\%) on the Chinese dataset, which is comparable to the previous Italian validation. Critically, emotionally valenced speech (both positive and negative) significantly outperformed neutral speech. This comparable performance between positive and negative tasks supports the emotional arousal hypothesis. Most importantly, EEG analysis revealed significant correlations between the model's speech-derived depression estimates and neural oscillatory patterns (theta and alpha bands), demonstrating alignment with established neural markers of emotional dysregulation in depression. This alignment, combined with the model's cross-linguistic robustness, not only supports that the CDMA framework's approach is a universally applicable and neurobiologically validated strategy but also establishes a novel paradigm for the neurophysiological validation of computational mental health models.
\end{abstract}


\section{Introduction}
\IEEEPARstart{A}{ccording} to the World Health Organization, depression affected approximately 3.8\% of the global population in 2023~\cite{WHO2023_Depression}. Traditional depression diagnosis predominantly relies on patient self-reporting combined with clinical observation and interviews. However, this conventional approach faces significant limitations, including patients' varying levels of self-awareness, willingness to disclose symptoms, and memory biases~\cite{hunt2003self,clement2015impact}, as well as the inherent subjectivity in clinicians' personal experience and judgment~\cite{mitchell2009clinical}. These factors collectively compromise diagnostic accuracy and reliability. The urgent need for objective, quantifiable biomarkers to supplement clinical assessment is therefore paramount for improving early screening and providing personalised patient care~\cite{li2025automated}.

Among potential biomarkers, speech, as the natural medium of communication in clinical interviews and the primary channel for expressing internal affective and cognitive states, has become a major focus for the development of automatic detection approaches~\cite{leal2024speech, niu2023wavdepressionnet, sun2025weakly, niu2025examining}. Speech-based depression detection typically employs two types of speech, i.e., \emph{read} speech and \emph{spontaneous} speech. Both have been demonstrated to contain complementary information relevant to depression identification~\cite{kiss2017comparison,alghowinem2013detecting}. Consequently, an increasing number of studies have begun exploring the fusion of these types of speech to enhance depression detection performance~\cite{ilias2024cross,tao2024cross}.

However, the promise of this fusion is undermined by an oversimplification in how spontaneous speech is handled. A dominant paradigm often treats the entirety of a participant's spontaneous speech as a monolithic block of data for analysis~\cite{aloshban2021you,alsarrani2022thin}. This analysis unavoidably flattens a complex emotional landscape, where speech representing different dimensional and even contradictory affective states is processed identically, despite compelling evidence that it is precisely these emotion-dependent patterns that serve as key indicators of a depressive state~\cite{zhao2022vocal,konig2022detecting}. Recognising this limitation, a more nuanced line of research has begun to investigate the role of different emotional contexts within spontaneous speech for depression detection~\cite{lin2025acoustic,teng2025enhanced}. However, the findings from this work are not entirely consistent. Some studies report that positive-valence tasks yield the highest classification accuracy, arguing that the blunted response to positive stimuli (anhedonia) creates the sharpest acoustic contrast~\cite{lin2025acoustic}. Conversely, a strong theoretical rationale suggests negative-valence tasks should be highly effective, as they are tailored to engage the specific cognitive impairments central to depression, such as the failure to inhibit negative information and associated ruminative processes~\cite{Cummins2015}. This apparent contradiction raises a more fundamental question about the relative importance of emotional \emph{valence} versus emotional \emph{arousal}. This leads to our central hypothesis that for depression detection, the intensity of an emotional state (arousal) is a more critical factor than its polarity (valence) in eliciting discriminative acoustic markers.

A second research gap impeding the clinical translation of these models is their cross-linguistic generalizability. While some research suggests that certain linguistic and acoustic features~\cite{tackman2019depression} and acoustic features of depression show similarities across languages~\cite{cummins2023multilingual}, these findings have been predominantly observed within related language families. These similarities have enabled reasonable cross-linguistic generalisation of depression detectors within European language systems~\cite{kiss2017investigation, costantini2022emotion}. However, this generalisability often fails when extending to typologically distant languages, particularly between tonal (e.g., Chinese) and non-tonal (e.g., English) systems, where systematic validation remains severely limited~\cite{lim2025lightweight}. 

% The most profound research gap is that even high-performing depression detectors lack interpretability, leaving their clinical trustworthiness in question without a clear understanding of their biological underpinnings. To bridge this gap, Electroencephalography (EEG) offers a direct, objective measure of the neural dynamics of emotional processing. A robust body of literature has used event-related potentials (ERPs) to delineate attentional biases in Major Depressive Disorder (MDD), revealing attenuated early sensory encoding of positive stimuli and hypervigilance toward negative content~\cite{dai2011deficient,zhao2015early}. Beyond phase-locked signals, non-phase-locked neural oscillations provide further insight; decreased alpha event-related desynchronization (ERD) is associated with heightened cortical activation and attentional engagement, while theta synchronization reflects cognitive control during emotional processing~\cite{li2022lack,li2023prioritizing,diao2017electroencephalographic}. However, no study has yet attempted to validate a speech-based depression model by directly correlating its predictions with these established neural markers of emotional dysregulation.

The most profound research gap is that even high-performing depression detectors lack interpretability, leaving their clinical trustworthiness in question without a clear understanding of their biological underpinnings. To bridge this gap, Electroencephalography (EEG) offers a direct, objective measure of the fast neural dynamics of emotional processing. A robust body of literature has used EEG to delineate attentional biases in Major Depressive Disorder (MDD), revealing attenuated early sensory encoding of positive stimuli and hypervigilance toward negative content~\cite{dai2011deficient,zhao2015early,li2022lack,li2023prioritizing,diao2017electroencephalographic}. However, no study has yet attempted to validate a speech-based depression model by directly correlating its predictions with these established neural markers of emotional dysregulation.

This study extends the Cross-Data Multilevel Attention (CDMA) framework~\cite{tao2024cross}, which was initially validated on Italian speech. Here, the primary contribution is not the proposal of a new model architecture but rather a rigorous investigation that aims to address the three fundamental gaps outlined above. Specifically, we apply this framework to a typologically distinct language dataset, MODMA Chinese Mandarin corpus~\cite{cai2022multi}, to make the following contributions:
\begin{itemize}
    \item we provide the first empirical evidence supporting the hypothesis that emotional arousal in speech, rather than valence, is a more critical factor for depression detection;
    \item we demonstrate the cross-linguistic generalizability of the CDMA framework by achieving state-of-the-art performance on a typologically distinct language;
    \item and we establish novel neurophysiological correlations between the model's depression estimates and corresponding EEG oscillatory activity, providing the first neurobiological validation for a computational speech-based approach to automatic depression detection.
\end{itemize}

The rest of this paper is organised as follows: Section~\ref{sec:related} reviews the relevant literature, Section~\ref{sec:app} details the methodology, Section~\ref{sec:res} presents the experimental findings, Section~\ref{sec:dis} discusses insights and Section~\ref{sec:con} draws conclusions.
\section{Related work}~\label{sec:related}
\subsection{Speech Types and Emotional Contexts in Automatic Depression Detection}
Early research on speech-based depression detection predominantly focused on a single speech type. For instance, \cite{tao2020spotting} utilised read speech to extract stable acoustic baselines, demonstrating that reading speed and pause duration are effective markers (reducing error rates by over 50\%). In contrast, \cite{mitra2015effects} focused on spontaneous speech, finding that modulation features extracted from spontaneous speech yielded a 25\% relative error reduction compared to read speech baselines. Recent approaches have advanced the processing of spontaneous speech through sophisticated architectures but often without differentiating emotional contexts. For instance, \cite{aloshban2021you} proposed a multimodal framework using Bidirectional Long Short-Term Memory networks that fused acoustic and linguistic features at the clause level; however, the model aggregated all clauses from the spontaneous speech into a single depression prediction regardless of the specific topic's valence. Similarly, \cite{alsarrani2022thin} introduced a ``thin slices" approach that improved detection performance by segmenting recordings into 128 blind slices, explicitly acknowledging in their conclusion that this ``blind segmentation" overlooked the variable emotional content within the spontaneous speech.

Empirical evidence from emotion-dependent speech pattern analysis challenges this aggregation strategy. For example, \cite{konig2022detecting} analysed speech from 118 participants and reported that individuals with higher depression scores produced significantly more words (increased verbosity), specifically during positive and negative story narrations but not in neutral contexts. In the acoustic domain, \cite{zhao2022vocal} applied multivariate analysis of variance on a Chinese clinical dataset and reported that Mel-frequency Cepstral Coefficients (MFCCs) (specifically MFCC2, MFCC3, and MFCC8) exhibited statistically significant variations across positive, neutral, and negative emotion tasks. Notably, they found that specific features, such as MFCC7 extracted from negative tasks, were highly predictive of Patient Health Questionnaire-9 (PHQ-9) scores ($\beta=0.90$), whereas neutral task features showed weaker correlations.

The divergence in identifying the optimal emotional context reflects distinct pathophysiological hypotheses. For instance, \cite{lin2025acoustic} reported superior classification performance in positive-valence tasks (87.5\% accuracy), attributing this to the ``positive attenuation" mechanism where patients fail to exhibit the expected acoustic buoyancy associated with joy. Conversely, \cite{Cummins2015} provided evidence favoring the diagnostic utility of negative-valence contexts, noting that the cognitive load induced by rumination and the failure to inhibit negative information lead to measurable prosodic delays and pauses. These conflicting pieces of evidence highlight that valence-specific markers may be highly task-dependent, lacking a unified framework to explain detection performance across the emotional spectrum.

\subsection{Cross-Linguistic Generalizability Challenges in Automatic Depression Detection}
Efforts to identify language-independent markers have yielded promising evidence regarding specific feature robustness. For instance, \cite{tackman2019depression} conducted a multi-lab synthesis and identified that first-person pronoun usage and negative emotion word frequency serve as consistent linguistic markers across diverse communicative contexts. Similarly, \cite{cummins2023multilingual} analysed remotely collected speech from English, Spanish, and Dutch cohorts, reporting that reduced speaking rate was a robust acoustic marker consistently associated with depression severity across these European languages.

Building on these feature-level similarities, model transferability has been successfully demonstrated within typologically similar language families. For instance, \cite{kiss2017investigation} employed Support Vector Regression (SVR) to predict depression severity across German, Hungarian, and Italian. Their experiments showed consistent performance, with Root Mean Square Error (RMSE) values ranging between 8.27 and 9.83 for cross-lingual tests, and an exceptionally low RMSE of 5.09 when training on German/Hungarian and testing on Italian. Similarly, \cite{costantini2022emotion} utilised Multi-Layer Perceptrons (MLP) on the Emofilm dataset, achieving over 80\% classification accuracy across English, Italian, and Spanish, suggesting that within Western languages, acoustic affective markers are robust to linguistic variation.

However, this generalizability breaks down when bridging the gap between tonal (e.g., Mandarin) and non-tonal (e.g., English) languages. For example, \cite{lim2025lightweight} proposed a lightweight multimodal architecture using MLP-Mixer for audio and Transformers for text. While the model achieved an F1-score of 81\% on the Chinese EATD-Corpus, the exact same architecture yielded an F1-score of only 67\% on the English DAIC-WOZ dataset, indicating a fundamental disparity in feature discriminability. This gap was exemplified in studies cited by \cite{qin2025cross}, where a model trained on English data plummeted to an accuracy of 48.72\% when tested on Chinese Mandarin data. These failures indicate that standard prosodic features (e.g., F0 variance) dominant in non-tonal languages are likely to fail to capture depression markers in tonal languages, where pitch variations are constrained by lexical semantic requirements.

% %
\subsection{Neural Dysfunction during Emotional Face Processing}
EEG studies have extensively utilised event-related potentials (ERPs) to delineate the attentional bias in MDD. ERPs capture distinct neural processing stages through specific components: P1 reflects early sensory encoding, N170 indicates structural face analysis, and P300 represents late cognitive evaluation. Patients with MDD exhibit attenuated P1 amplitudes over occipital regions specifically for positive stimulus, reflecting impaired initial engagement with positive stimuli; this deficit correlates with anhedonia severity~\cite{dai2011deficient}. Conversely, enhanced N170 amplitudes to sad faces at temporo-occipital sites indicate hypervigilance toward negative emotional content in MDD~\cite{zhao2015early}. During late-stage processing (300–500 ms), increased P300 amplitudes for sad faces in parietal regions have been reported, indicating deficient cognitive inhibition and excessive facilitation~\cite{dai2011deficient}.

Apart from phase-locked ERP signals, ongoing non-phase-locked oscillatory signals could also provide more evidence for the neural impairments in MDD. Alpha rhythms (8–12 Hz) are highly related to selective attention; a decrease in alpha power, known as alpha event-related desynchronization (ERD), is associated with increased cortical activation indicative of attentional engagement~\cite{li2021visual,li2022lack,li2023prioritizing}. Previous work demonstrated elevated alpha ERD in posterior regions during emotional face viewing~\cite{sollfrank2021effects}. Conversely, theta synchronization (3–7 Hz) intensifies during negative face exposure, facilitating sustained negative information processing~\cite{diao2017electroencephalographic}. However, few studies focused on the temporal dynamics of neural oscillations associated with face processing in MDD. Whether patients with MDD exhibit cortical hyperactivation and maladaptive inhibition of negative face remains unclear. To address it, we mainly focused on alpha and theta oscillations during face processing with different emotions to figure out whether these oscillatory signals could serve as markers for the neural dysfunction observed in MDD.
\section{The Approach}\label{sec:app}
% Our proposed approach consists of four main steps: \emph{feature extraction}, \emph{segmentation}, \emph{detection}, and \emph{aggregation}, as outlined in Figure~\ref{scheme}.
%
%
\subsection{The Data}
\subsubsection{Participants}
The MODMA dataset was collected under approved ethical guidelines as detailed by its original creators, and all participants provided informed consent~\cite{cai2022multi}. 
For speech-based depression detection, we utilised participants from the recordings of the spoken language experiment subset in MODMA, which originally included 52 participants (23 MDD patients: 16 males and 7 females, aged 16-56 years; 29 healthy controls: 20 males and 9 females, aged 18-55 years). 
For EEG analysis, a total of 17 participants with MDD (11 males and 6 females, 31.76 $\pm$ 10.94 years) and 21 Healthy Controls (HCs) (14 males and 7 females, 32.19 $\pm$ 10.18 years) who had both EEG and speech recordings were selected for further analysis. During EEG preprocessing, three participants (1 HC and 2 MDDs) were excluded due to excessive artifacts resulting in fewer than 50\% valid EEG trials. In the time-frequency analysis stage, 2 additional MDD participants were excluded due to power values deviating more than three standard deviations from the group mean. As a result, 13 MDD patients (8 males and 5 females, 32.08 $\pm$ 10.36 years) and 20 HCs (14 males and 6 females, 32.10 $\pm$ 10.43 years) were retained for final analysis.

\subsubsection{Data Acquisition}
For this work, we utilised reading recordings and interview responses covering positive, neutral, and negative emotions to evaluate the effectiveness of our proposed approach. The reading task (recording 19) included a standard narrative passage (``The North Wind and the Sun''). Interview recordings consisted of 18 questions derived from clinical depression assessment scales, with 6 questions each targeting positive (recordings 1-6), neutral (recordings 7-12), and negative (recordings 13-18) emotional valences. For EEG, participants performed an emotion-modulated dot-probe task using grayscale facial images from the Chinese Facial Affective Picture System (CFAPS)~\cite{lu2005development}, presented in emotional-neutral pairs (happy, sad, or fear vs. neutral), see~\cite{cai2022multi} for more details.
\subsection{Feature Extraction and Segmentation}
To ensure methodological consistency with the original CDMA study~\cite{tao2024cross}, we adopted identical feature extraction protocols using the OpenSMILE toolkit~\cite{eyben2013recent} with configurations from the Androids corpus~\cite{tao2023androids}. We extracted 16 acoustic descriptors (energy, MFCCs, fundamental frequency, zero-crossing rate, and voicing probability) along with their first-order temporal derivatives, yielding 32-dimensional feature vectors ($D=32$) for each temporal frame. Feature vectors $\vec{x}_u$ were computed using 25 ms sliding windows with 10 ms frame shifts. This approach ensures a direct and fair comparison of the model's performance across different languages (Italian and Mandarin Chinese). Moreover, this established feature set has been widely validated for capturing key prosodic, spectral, and voice quality information relevant to depression~\cite{Cummins2015}.

\begin{figure*}[h]
  \centering
  \includegraphics[width=\textwidth]{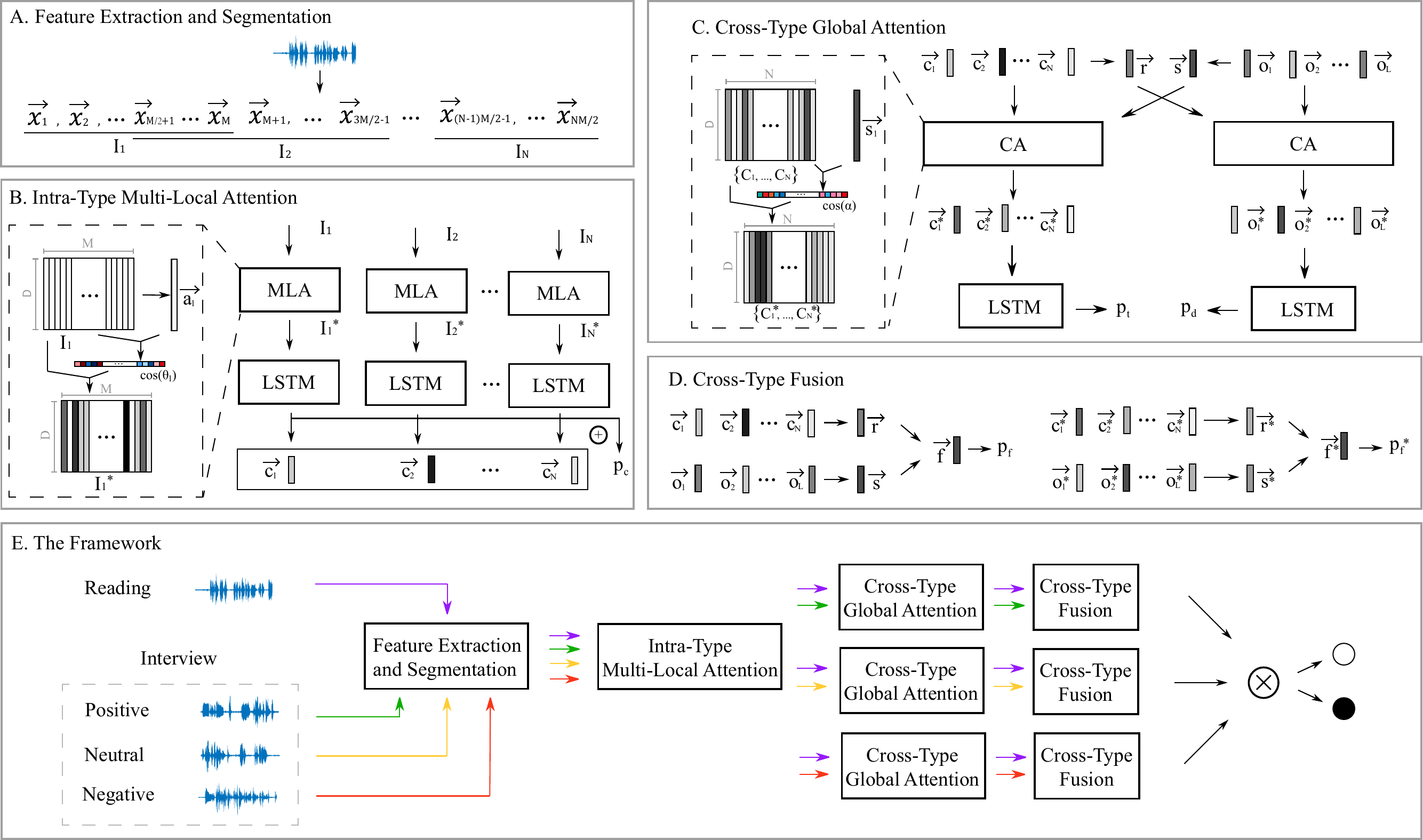}
  \caption{The figure illustrates the proposed framework. Purple arrow shows read speech processing; green, yellow, and red arrows show positive, neutral, and negative spontaneous speech processing, respectively. The symbol $\otimes$ indicates majority voting that combines classification results from all speech types to produce the final depression prediction.}\label{scheme}
\end{figure*}
As illustrated in Fig.~\ref{scheme}A, this processing produces feature sequences $X = \{\vec{x}_1, \ldots, \vec{x}_T\}$ for read speech. Although Fig.~\ref{scheme}A illustrates using read speech as an example, the same methodology is applied uniformly to spontaneous speech. We use $Z = {\vec{z}_1, \ldots, \vec{z}_Q}$ to represent one category of spontaneous speech (either negative, neutral, or positive interview response). Note that the same representation applies uniformly across all three emotional categories throughout the entire approach. Sequence lengths $T$ and $Q$ typically differ due to varying recording durations.

% \subsection{Segmentation}
Following feature extraction, each recording becomes a high-dimensional time series of acoustic feature vectors. Since recordings span several seconds to minutes, the resulting sequences contain thousands of feature vectors, which are too long for direct neural model processing. To maintain consistency with established protocols~\cite{tao2023androids}, we partition sequences $X$ and $Z$ into overlapping temporal segments of $M=128$ consecutive feature vectors using a sliding window with stride $M/2$, providing 50\% overlap for temporal continuity.

For read speech sequences, this segmentation generates a series of segments $I = \{I_k\}$ ($k \in [1, N]$), where each segment $I_k$ contains $M$ feature vectors and $N$ denotes the total number of segments.  For spontaneous speech, since each emotional category contains 6 interview responses, segmentation is performed at the individual response level to preserve discourse structure. This response-based approach produces a set of segments $V = \{V_m\}$ ($m \in [1, L]$), where $L$ represents the total number of frames, with the constraint that no segment spans across different responses within the same emotional category. This procedure yields segmented read speech ($I$) and segmented spontaneous speech ($V$) for each emotional category, with consistent representations across all three emotional states.
% The segmentation yields two distinct sets of temporal segment sequences: $I$ for read speech and $V$ for one emotional category of spontaneous speech, with consistent segment-level representations across all three emotional states (negative, neutral, positive).
%
\subsection{Detection: Cross-Data Multilevel Attention}
\subsubsection{Intra-Type Multi-Local Attention}\label{it-mla}
Following segmentation, the sequences $I$ and $V$ represent read and spontaneous speech as collections of temporal segments, respectively. The first detection stage, termed \emph{Intra-Type Multi-Local Attention} (IT-MLA), processes each speech type independently to identify and emphasise localised segment-level acoustic patterns that are indicative of depression. Fig.~\ref{scheme}B depicts the IT-MLA detection stage for read speech, the identical procedure is applied to spontaneous speech.

Within each segment $k$, we compute the average feature vector $\vec{a}_k$ as a local contextual reference. Each feature vector $\vec{x}_i$ within the segment is then adaptively weighted according to its acoustic alignment with $\vec{a}_k$, using a cosine similarity-based attention mechanism:
\begin{align}
    &\vec x_{i}^{*} = \vec x_i +\frac{exp(\cos\theta_i)}{ {\textstyle \sum_{j=1}^{M}} exp(\cos\theta_j)}\cdot\vec x_i, \\
    &\cos\theta_i=\vec a_k\vec x_i/||\vec a_k||\cdot ||\vec x_i||,
\end{align}
where $i \in \{1,\ldots,M\}$, and $\cos\theta_i$ represents the cosine similarity between vector $\vec{x}_i$ and segment-average vector $\vec{a}_k$. This attention mechanism selectively amplifies feature vectors with stronger alignment to the local acoustic context, enhancing discriminative capacity of depression-related patterns within each segment.

The attention-enhanced segments are subsequently processed through a Long Short-Term Memory network (LSTM)~\cite{hochreiter1997long} for binary classification at the segment level. Speaker-level depression likelihood is estimated as the proportion of segments classified as depressive. This generates two probability estimates: $p_c$ from read speech and $p_o$ from spontaneous speech, with the same framework applied across all three emotional states.
\subsubsection{Cross-Type Global Attention}\label{ct-ga}
The second component, termed \emph{Cross-Type Global Attention} (CT-GA), leverages inter-type information exchange between speech types while highlighting shared discriminative patterns present in both read and spontaneous speech. This cross-type approach enables each speech type to inform and enhance the representation of the other, capturing complementary depression-related cues across different speaking contexts.

Following IT-MLA, the LSTM networks generate hidden state representations for each temporal segment. These hidden states are temporally averaged to produce segment-level representations $\vec{c}_k$ for read speech, as depicted in Fig.~\ref{scheme}B. Similarly, we use $\vec{o}_k$ to denote the corresponding average segment-level representations obtained through spontaneous speech. These vectors form the sequences $C$ and $O$, respectively, as illustrated in Fig.~\ref{scheme}C.

The CT-GA mechanism applies cross-type attention to transform both sequences. We present the transformation for $\vec{c}_k$, with the analogous operation applied to $\vec{o}_k$:
\begin{align}
    &\vec c_{k}^{*} = \vec c_k +\frac{exp(\cos\alpha_k)}{ {\textstyle \sum_{j=1}^{N}} exp(\cos\alpha_j)}\cdot\vec c_k, \\
    &\cos\alpha_k=\vec s \vec c_k/||\vec s||||\vec c_k||, \label{fom}
    %&\vec o_{m}^{*} = \vec o_m +\frac{exp(\cos\beta(m))}{ {\textstyle \sum_{m=1}^{L}} exp(\cos\beta(m))}\cdot\vec o_m, \\
    %&\cos\beta_m=\vec r \vec o_m/||\vec r||||\vec o_m||,
\end{align}
where $\vec{s}$ represents the global average of spontaneous speech sequence $O$, serving as the cross-type reference for adapting read speech representations. Conversely, for spontaneous speech transformation, $\vec{r}$ (the global average of sequence $C$) serves as the guiding reference. This bidirectional attention mechanism enhances features exhibiting consistent patterns across both speech types, amplifying depression-relevant information.

The attention-enhanced sequences $C^*$ and $O^*$ are fed into separate LSTM networks, generating sequences of refined hidden states. These are temporally averaged to produce speaker-level representations, then processed through a softmax layer to yield depression probability estimates $p_t$ and $p_d$, corresponding to read-speech-informed and spontaneous-speech-informed predictions, respectively.
\subsubsection{Cross-Type Fusion}\label{ctf}
Following CT-GA, four distinct sequences emerge: $C$, $O$, $C^*$, and $O^*$, with corresponding global averages $\vec{r}$, $\vec{s}$, $\vec{r}^*$, and $\vec{s}^*$, as illustrated in Fig.~\ref{scheme}D. This enables construction of two composite vectors representing different processing stages: $\vec{f} = \vec{r} + \vec{s}$ and $\vec{f}^* = \vec{r}^* + \vec{s}^*$. Vector $\vec{f}$ encapsulates information from the intermediate stage following IT-MLA, while $\vec{f}^*$ captures enhanced representations from CT-GA. Both vectors are independently processed through separate softmax layers, yielding depression probability estimates $p_{f}$ and $p_{f^*}$. Since vectors $\vec{f}$ and $\vec{f}^*$ integrate averaged representations from distinct speech types, this stage is designated as \emph{Cross-Type Fusion} (CTF).
\subsubsection{Aggregation}
The framework generates six probability estimates throughout the detection pipeline, combined through averaging:
\begin{equation}
\hat p = \frac{1}{|\mathcal{B}|}\sum_{v\in\mathcal B} p_v,
\end{equation}
where $\mathcal{B} = \{c, o, f, f^*, t, d\}$ represents the set of all probability estimates. The final classification assigns a speaker to the depressed category when $\hat{p} > 0.5$. For training, all probability estimates are simultaneously optimised using a unified loss function:
\begin{equation}
    \mathcal{L}=-\frac{1}{K}\sum_{v\in\mathcal B}{[y\log(p_v) + (1 - y)\log(1 - p_v)]},
\end{equation}
where $K$ is the total number of training samples, and $y$ is the ground truth label.

Since the framework processes each emotional category of spontaneous speech separately with read speech, three independent predictions are generated for negative, neutral, and positive spontaneous speech processing. The final speaker-level depression prediction combines these three predictions through majority vote, as shown in Fig.~\ref{scheme}E.
\subsection{Experimental setup and training}
All experiments were conducted using 5-fold cross-validation. To ensure \emph{person-independent} evaluation, all recordings from each speaker were assigned to the same fold, guaranteeing no participant appears in both training and testing sets. This design ensures models learn depression-related patterns rather than speaker-specific characteristics.
The LSTM networks used 32 hidden units, with training at a learning rate of $10^{-3}$, batch size 32, over 50 epochs. We employed PyTorch's default weight initialisation with RMSProp optimizer~\cite{tieleman2012lecture}. To account for stochastic weight initialisation variability, each configuration was executed 50 times independently. All performance metrics are presented as mean values with standard deviations across these 50 iterations. 
\subsection{EEG Methods}
\subsubsection{Preprocessing}
EEG preprocessing was performed using EEGLAB v2020.0 and MATLAB R2023b. Raw data were re-referenced to the average reference, bandpass filtered (0.1-30 Hz), notch filtered (48-52 Hz) to remove line noise, and downsampled to 250 Hz. Epochs were extracted from -1 to 2 s relative to stimulus onset, with the interval from -0.5 to 1.5 s used for artifact detection. Independent Component Analysis (ICA) with Principal Component Analysis (PCA) dimensionality reduction (30 components) was applied to identify and remove artifacts. Epochs with voltage fluctuations exceeding $\pm$\SI{100}{\micro\volt} or containing residual electrooculographic (EOG) artifacts were excluded from further analysis.
\subsubsection{Time-Frequency Analysis}
Event-Related Spectral Perturbation (ERSP) analysis was conducted using Short-Time Fourier Transform-based methods to extract time-frequency representations from 1-30 Hz with 1.25 Hz frequency resolution using EEGLAB's \texttt{pop\_newtimef} function. EEG Baseline correction was applied using the pre-stimulus interval from -300 to -100 ms. Spectral power in the theta (3-7 Hz) and alpha (8-12 Hz) frequency bands was averaged across time and frequency bins for statistical comparison between groups.
\subsubsection{Group-Level Statistical Analysis}
Two-tailed independent-sample $t$-tests were conducted to compare spectral power between MDD patients and HCs across predefined regions of interest (ROIs): frontal (13 electrodes) and parieto-occipital (17 electrodes) regions. Statistical comparisons were performed within specific time windows: early (0-200 ms) and late (400-600 ms) time windows.
\subsubsection{Correlation Analysis Between Model Logits and Neural Features}
Spearman correlation analysis was performed to assess the relationship between mean ERSP power and model-derived logits obtained from the CDMA framework. Correlations were computed separately for three emotional spontaneous conditions (positive, neutral, negative) and differential emotion scores ($\Delta$positive = positive logits minus neutral logits; $\Delta$negative = negative logits minus neutral logits) within theta and alpha frequency bands, across the defined frontal and parieto-occipital ROIs and the specified early and late time windows. 
\section{Experiments and results}\label{sec:res}
\subsection{CDMA Peformance Analysis}
We next evaluate the performance of the full framework described in Section~\ref{sec:app}. A comprehensive ablation study demonstrating the necessity of each component within the core CDMA framework was conducted in the study that originally proposed the framework~\cite{tao2024cross}. This allows the current analysis to further investigate the framework's detection performance across different linguistic and emotional contexts, validating these findings neurophysiologically with EEG data.

\subsubsection{Comparisons Between Emotional Spontaneous Speech}
\begin{table}[t!]
\caption{Performance of the proposed framework in terms of accuracy (Acc.), precision (Prec.), recall (Rec.), and F1-score for models combining read speech with positive, neutral, and negative spontaneous speech, respectively. Metrics are presented as mean ± standard deviation across 50 iterations.}\label{tab:performance}
\centering
\begin{tabular}{ccccc}
\hline
         & Acc.(\%)     & Prec.(\%)    & Rec.(\%)      & F1(\%)    \\ \hline
Read + Positive & 83.8$\pm$2.6 & 83.0$\pm$4.4 & 89.2$\pm$4.7 & 84.4$\pm$2.5 \\
Read + Neutral  & 81.5$\pm$2.1 & 79.3$\pm$3.6 & 87.9$\pm$5.6 & 82.1$\pm$2.6 \\
Read + Negative & 84.0$\pm$2.1 & 82.4$\pm$3.9 & 89.8$\pm$4.0 & 85.0$\pm$2.2 \\ \hline
\end{tabular}
\end{table}
To evaluate the statistical significance of performance differences between different models, we conducted two-tailed $t$-tests on the F1-scores obtained from 50 experimental iterations. Specifically, we compared the performance of models that combine read speech with different emotional categories of spontaneous speech (negative, neutral, positive), as shown in Table~\ref{tab:performance}. 

The combination of read speech with positive, neutral, and negative spontaneous speech yielded average F1-scores of 84.4\%, 82.1\%, and 85.0\%, respectively. Statistical analysis suggested that combining read speech with either positive or negative spontaneous speech significantly outperformed the combination with neutral spontaneous speech according to two-tailed $t$-tests with all $p < 0.001$. These results demonstrate that when integrated with read speech, emotionally spontaneous speech (both positive and negative) provides more discriminative acoustic patterns for depression detection compared to neutral spontaneous speech. 

% Notably, no statistically significant difference was observed between the combination of read speech with positive spontaneous speech and the combination of read speech with negative spontaneous speech ($p > 0.05$). This lack of significant difference suggests that when combined with read speech, emotional arousal in spontaneous speech, rather than valence polarity, may be the critical factor in enhancing depression detection performance. This observation has important implications for developing depression detectors, indicating that any emotionally engaging conversational context can effectively complement read speech to achieve robust depression detection.
% Notably, no statistically significant difference was observed between the positive task (Read + Positive) and the negative task (Read + Negative) (p>0.05). These findings provide strong empirical evidence for the arousal hypothesis. 
% The observation that both high-arousal tasks significantly outperformed the low-arousal (Read + Neutral) task suggests that emotional engagement is crucial for eliciting discriminative speech markers. 
% Furthermore, the comparable performance between the positive and negative tasks, despite their opposite valences, indicates that the intensity of the emotional state (arousal) is a more critical factor for detection accuracy than the polarity of the emotion (valence). This has important implications for developing depression detectors, suggesting that any emotionally engaging context can effectively complement read speech to achieve robust detection.
Notably, no statistically significant difference was observed between the positive task (Read + Positive) and the negative task (Read + Negative) ($p>0.05$). The performances between the positive and negative tasks are comparable, despite their opposite valences. This indicates that the intensity of the emotional state (arousal) is a more critical factor for detection accuracy than the polarity of the emotion (valence). These findings provide strong empirical evidence for the arousal hypothesis. This has important implications for developing depression detectors, suggesting that any emotionally engaging context can effectively complement read speech to achieve robust detection.
\subsubsection{Comparisons to state-of-the-arts}
\begin{table}[t!]
\caption{Best performance of the proposed framework compared with baseline and state-of-the-art (SOTA) methods on the MODMA dataset. The table shows the highest F1-score and corresponding accuracy, precision, and recall for models combining read speech with positive, neutral, and negative spontaneous speech individually, and the combined result using majority voting across all three emotional categories. Results for RNN~\cite{rejaibi2022mfcc} and DepAudioNet~\cite{ma2016depaudionet} are as reported in~\cite{du2023depression}, while those for DMPF, Vlad-GRU, ConvNeXt, and TimesNet are as reported in~\cite{zhao2025decoupled}. The symbol `-' indicates that the metric was not reported in the original study.}\label{tab:best_performance}
\centering
\begin{tabular}{ccccc}
\hline
                            & Acc.(\%)     & Prec.(\%)    & Rec.(\%)      & F1(\%)    \\ \hline
DepAudioNet~\cite{ma2016depaudionet}   & - & - & - & 79.0 \\
Decision Tree~\cite{chen2021convenient}   & 83.4 & 81.9 & 79.0 & 80.5 \\
RNN~\cite{rejaibi2022mfcc}   & - & - & - & 70.0 \\
MLP~\cite{gheorghe2023using}    & 84.2 & 85.3 & 83.8 & 84.1  \\
CNN~\cite{gheorghe2023using}    & 84.2 & 85.3 & 83.8 & 84.0  \\ 
MSCDR~\cite{du2023depression}     & 85.7 & 79.2 & 90.5 & 85.6 \\
DCAN~\cite{qin2025cross}     & 81.1 & - & - & 78.4 \\
DMPF~\cite{zhao2025decoupled}& 76.9 & 76.8 & 76.4 & 76.5 \\
Vlad-GRU~\cite{shen2022automatic}& 63.9 & 66.7 & 55.3 & 60.4 \\
ConvNeXt~\cite{liu2022convnet}& 68.1 & 69.1 & 66.7 & 63.8 \\
TimesNet~\cite{wu2022timesnet}& 65.4 & 67.1 & 63.7 & 61.8 \\\hline
Read + Positive             & \textbf{88.7} & \textbf{89.2} & 93.3 & \textbf{89.6} \\
Read + Neutral              & 86.5 & 86.8 & 90.0 & 86.8 \\
Read + Negative             & 86.7 & 85.7 & 92.7 & 88.7 \\
Combination                 & 88.5 & 81.5 & \textbf{95.7} & 88.0 \\\hline
\end{tabular}
\end{table}
To contextualise our findings, Table~\ref{tab:best_performance} presents a comprehensive comparison between our proposed framework and a range of baseline and SOTA approaches on the MODMA dataset with person-independent split protocols. The results unequivocally demonstrate that our approach achieves superior performance across all evaluation metrics. Our method consistently outperforms a wide spectrum of models, from classic machine learning algorithms like Decision Trees to established deep learning architectures such as Recurrent Neural Networks (RNNs), Convolutional Neural Networks (CNNs), and dedicated models like DepAudioNet. Notably, our best-performing individual model (read speech with positive spontaneous speech) achieves an F1-score of 89.6\%, establishing a substantial margin of 4.0\% over the strongest SOTA competitor, MSCDR~\cite{du2023depression}. 

The combination result, obtained through majority voting across all three emotional categories, achieves an F1-score of 88.0\%, which does not exceed the performance of the individual best model. This suggests that the majority voting mechanism may introduce some conservative bias, as it requires consensus among multiple models that may have different strengths and weaknesses. However, the combination approach demonstrates a crucial advantage in clinical applications: it achieves the highest recall of 95.7\%, substantially exceeding both the individual models and state-of-the-art methods. This exceptional recall performance indicates that the ensemble approach minimises false negatives, ensuring that fewer individuals with depression are misclassified as healthy. 
Such high sensitivity is particularly valuable in depression screening scenarios, where missing positive cases could delay critical interventions and have serious clinical consequences.

Notably, the CDMA framework was originally developed and validated on Italian speech~\cite{tao2024cross}. The successful application to the typologically distinct Chinese Mandarin corpus, achieving comparable performance, provides compelling evidence for the cross-linguistic robustness of the proposed cross-type attention mechanisms. This consistent effectiveness suggests that the framework captures underlying depression-related acoustic patterns that may be largely language-independent, potentially representing universal indicators of emotional dysregulation tied to the physiological underpinnings of depression.
\subsection{EEG analysis}
To validate that the acoustic patterns identified by the speech model are associated with depression-related emotional dysregulation, we first examined EEG differences between MDD patients and HCs during emotional processing, then explored associations between these neural oscillatory patterns and our CDMA-derived depression estimates.
% We further analysed EEG differences between patients with MDD and HCs, and explored how these neural differences were associated with the CDMA speech model.
%
\subsubsection{Group Differences in Time-Frequency Representations}
\begin{figure}[h!]
 % \centering}
 \includegraphics[width=\columnwidth]{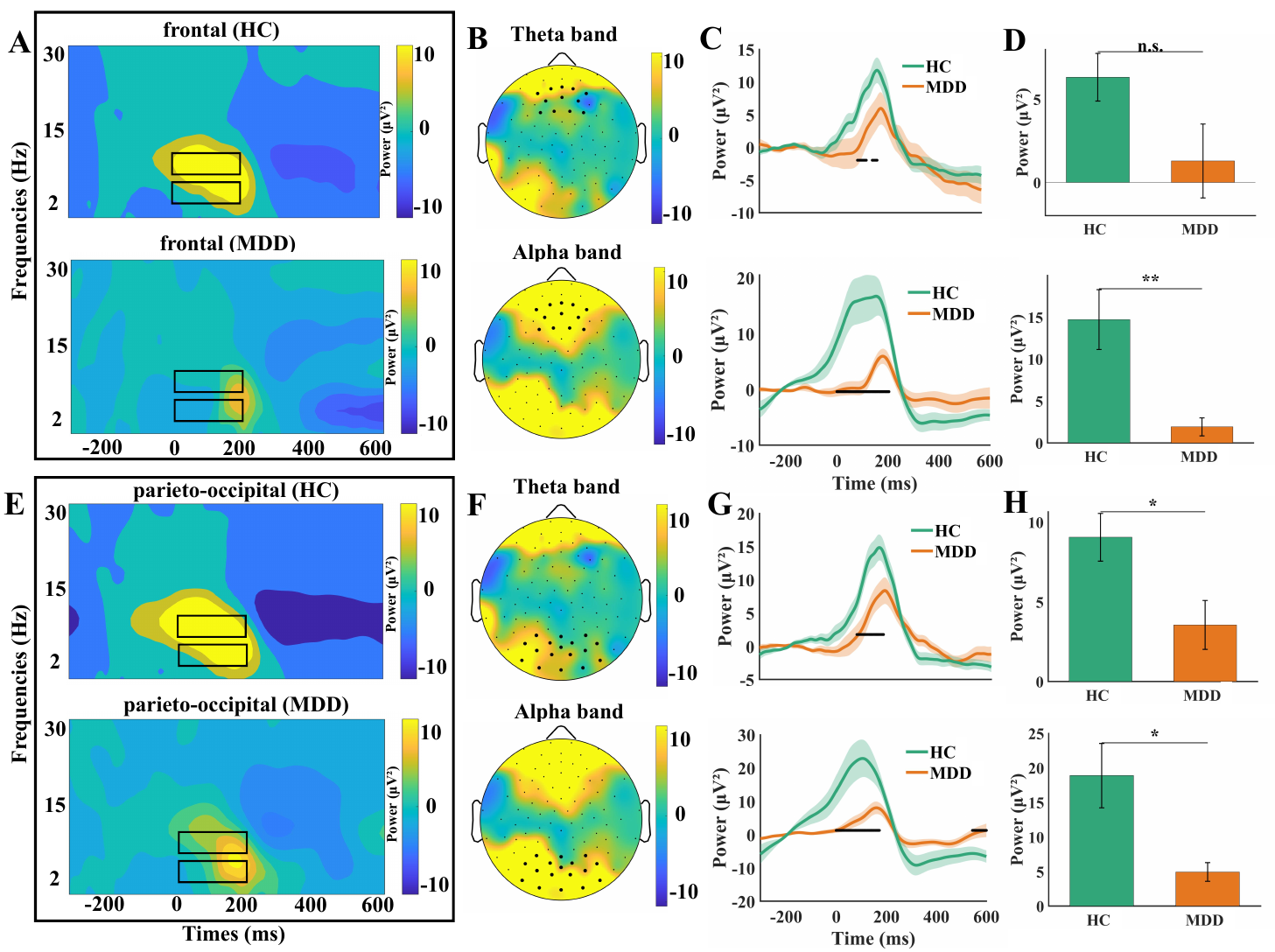}
 \caption{
 Time-frequency representations for the fearful facial expression condition. (A-D) Frontal region (13 electrodes including E23, E18, E16, E10, E3, E19, E11, E4, E20, E12, E5, E118): (A) Time-frequency representations of HCs and MDDs; (B) topographical distribution of the difference in theta (3-7 Hz) and alpha (8-12 Hz) band power between HCs and MDDs during 0-200 ms time windows; (C) time course of theta and alpha power with significant differences ($p < 0.05$); (D) bar plots of mean theta and alpha power within 0-200 ms; ** denotes $p < 0.01$. (E-H) Parieto-occipital region (17 electrodes, E62, E60, E67, E72, E77, E85, E59, E66, E71, E76, E84, E91, E65, E70, E75, E83, E90 included): (E)Time-frequency representations of HCs and MDDs; (F) topographical difference maps (HCs minus MDDs); (G) time course with significant time ($p < 0.05$); (H) bar plots of mean theta and alpha power within 0-200 ms. * denotes $p < 0.05$. Significant group differences are observed in both theta and alpha power for both regions.
  % Time-frequency analysis for the fear facial expression condition in HCs and MDD patients.
 }\label{fig:fear}
\end{figure}
\begin{figure}[t!]
 % \centering
 \includegraphics[width=\columnwidth]{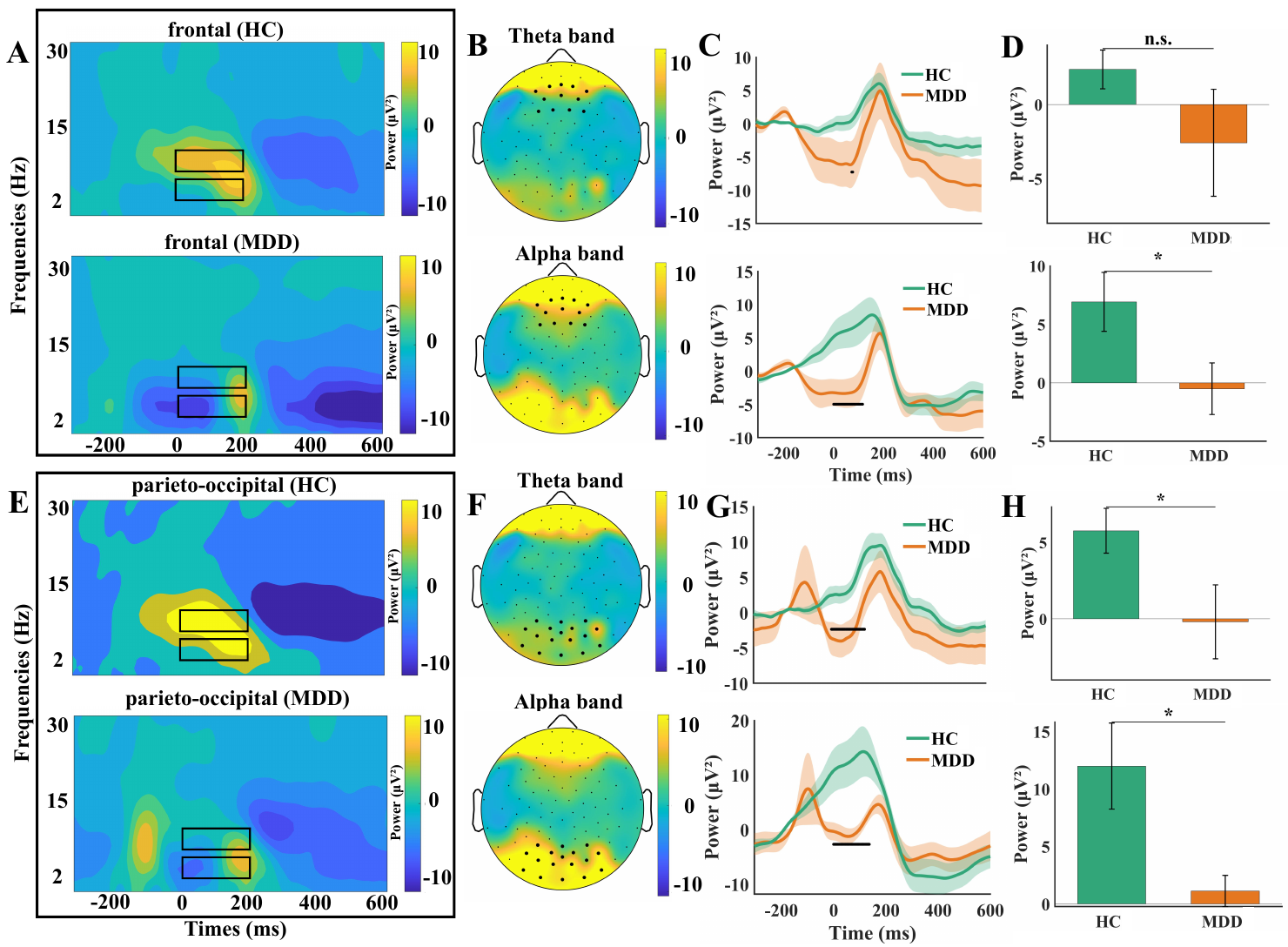}
 \caption{Time-frequency representations for the sad facial expression condition. (A-D) Frontal region: (A) Time-frequency power fluctuations in HCs and MDDs following stimulus onset; (B) topographical distribution of the difference in theta and alpha power between HCs and MDDs during the 0-200 ms time window, with stronger power observed in HCs; (C) time course of theta and alpha power with shaded areas indicating standard error of the mean (SEM); significant intervals in alpha are marked ($p < 0.05$). * denotes $p < 0.05$; (D) bar plots of mean theta and alpha power within 0-200 ms, showing a significant group difference in alpha ($p < 0.05$), while theta remains non-significant. (E-H) Parieto-occipital region: (E) Time-frequency representations of theta and alpha activity in HCs and MDDs; (F) topographical distribution of group differences (HCs minus MDDs), with higher theta and alpha power in HCs; (G) time course of theta and alpha power with significant intervals indicated ($p < 0.05$); (H) bar plots of mean theta and alpha power within 0-200 ms, with both frequency bands showing significant group differences. 
  % Time-frequency analysis for the sad facial expression condition in HCs and MDD patients.
}\label{fig:sad}
\end{figure}
% Time-frequency representations exhibited strong fluctuations within the theta and alpha frequency bands across various time windows following presentation of emotional faces. We found significant power differences were observed between MDD and HCs when presented with fearful (Fig.~\ref{fig:fear}) and sad (Fig.~\ref{fig:sad}) faces, as opposed to happy (Fig.~\ref {fig:happy}) faces.
Time-frequency analysis revealed oscillatory changes in theta and alpha frequency bands following presentation of emotional faces in both MDD and HC groups. Notably, significant group differences in oscillatory dynamics were mainly observed for negative emotional faces (fearful and sad) but not for positive faces (happy).

For negative facial expressions (fear and sad), time-frequency analysis revealed marked group differences in theta and alpha bands during the early time window (0-200 ms) following stimulus onset, shown as Fig.~\ref{fig:fear}A and Fig.~\ref{fig:sad}A. Topographical difference maps (HCs minus MDD), shown as Fig.~\ref{fig:fear}B and Fig.~\ref{fig:sad}B, demonstrated greater frontal theta and alpha activity in HCs compared to MDD patients. Similarly, the parieto-occipital region showed pronounced group differences in both frequency bands during this early window (Fig.~\ref{fig:fear}E-F and Fig.~\ref{fig:sad}E-F). Temporal dynamics of averaged power at frontal (Fig.~\ref{fig:fear}C and Fig.~ \ref{fig:sad}C) and parieto-occipital electrodes (Fig.~\ref{fig:fear}G and Fig.~\ref{fig:sad}G) confirmed significant group differences in the early window through independent-sample $t$-tests conducted at each time point ($p<0.05$).
Statistical analysis revealed non-significant frontal theta differences (all $p>0.05$) but significant differences in frontal alpha power (fear: $t(31) = 2.822$, $p = 0.008$, Cohen's d $= 1.005$; sad: $t(31) = 2.062$, $p = 0.048$, Cohen's d $= 0.735$) (Fig.~\ref{fig:fear}D and Fig.~\ref{fig:sad}D), while MDD patients showed significantly larger parieto-occipital ERDs in both theta (fear: $t(31) = 2.459$, $p = 0.020$, Cohen's d $= 0.876$; sad: $t(31) = 2.238$, $p = 0.033$, Cohen's d $= 0.797$) and alpha bands (fear: $t(31) = 2.364$, $p = 0.025$, Cohen's d $= 0.842$; sad: $t(31) = 2.265$, $p = 0.031$, Cohen's d $= 0.807$) (Fig.~\ref{fig:fear}H and Fig.~\ref{fig:sad}H).
These findings suggest that the reduced alpha power observed in MDD during early processing of negative faces reflects maladaptive sustained attention toward negative stimuli, while the attenuated theta power indicates deficient top-down cognitive control and impaired emotional regulation mechanisms~\cite{villafaina2019influence,disner2011neural}. 

\begin{figure}[t!]
 % \centering
 \includegraphics[width=\columnwidth]{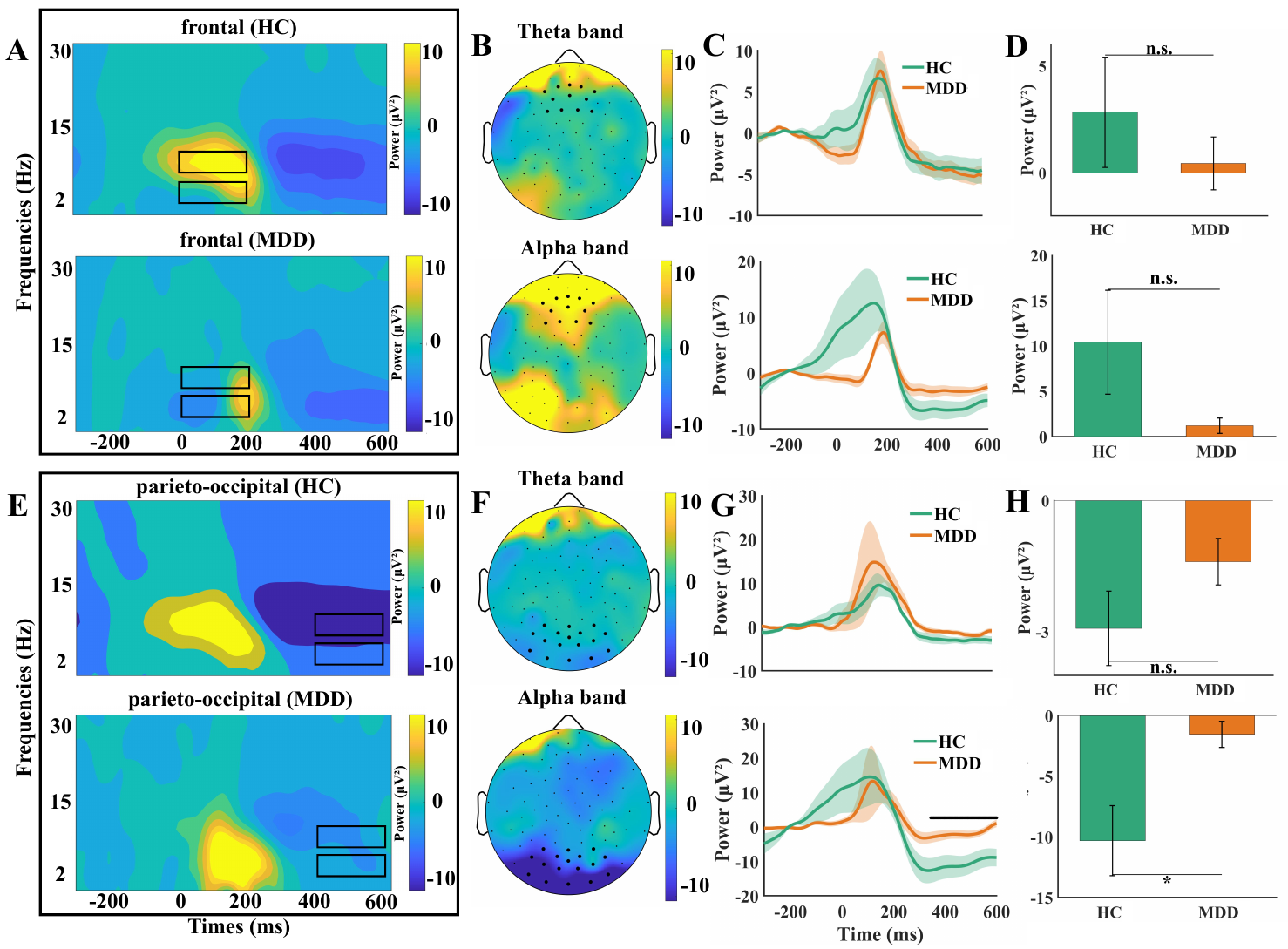}
 \caption{
 Time-frequency representations for the happy facial expression condition. (A-D) Frontal region: (A) Time-frequency representations in HCs and MDDs; (B) topographical distribution of the difference in theta and alpha band power between HCs and MDDs during the 0-200 ms time window, showing increased alpha power in HCs; (C) time course of theta and alpha power with shaded areas indicating standard error of the mean (SEM); no significant group differences were observed; (D) bar plots of mean theta and alpha power within 0-200 ms, confirming non-significant effects. (E-H) Parieto-occipital region: (E) Time-frequency representations of HCs and MDDs showing activity differences in the 400-600 ms window; (F) topographical distribution of group differences (HCs minus MDDs), showing higher theta and alpha power in MDDs; (G) time course of theta and alpha power with significant intervals in alpha indicated ($p < 0.05$). * denotes $p < 0.05$; (H) bar plots of mean theta and alpha power within 400-600 ms; alpha power shows a significant group difference ($p < 0.05$), while theta remains non-significant.
 % Time-frequency analysis for the happy facial expression condition in HCs and MDD patients.
 }\label{fig:happy}
\end{figure}
For positive facial expressions (happy), time-frequency analysis revealed different temporal and regional patterns compared to negative faces, with frontal regions showing considerable theta and alpha band activity changes within the early window (Fig~\ref{fig:happy}A) but parieto-occipital regions displaying later power differences between groups (Fig.~\ref{fig:happy}E). Topographical difference maps (Fig.~\ref{fig:happy}B and Fig.~\ref{fig:happy}F) indicated comparable frontal theta power between groups but lower frontal alpha power in MDD patients, while the parieto-occipital region showed greater theta and alpha activity in MDD patients during later processing. 
Temporal dynamics (Fig.\ref{fig:happy}C and Fig.\ref{fig:happy}G) revealed non-significant group differences for frontal alpha and theta power, as well as parieto-occipital theta power, while only parieto-occipital alpha power showed significant differences from 344 to 600 ms.
Statistical analysis revealed no significant frontal differences in either theta or alpha power during early processing, with all $p>0.05$ (Fig.~\ref{fig:happy}D), while MDD patients showed significantly greater parieto-occipital alpha power during later processing ($t(31) = -2.362$, $p = 0.025$, Cohen's d $= -0.841$) but non-significant theta differences ($p>0.05$), shown as Fig.~\ref{fig:happy}H.
While MDD patients initially process positive emotional stimuli similarly to HCs, as evidenced by the absence of group differences in early theta and alpha power, elevated alpha power during the later time window suggests decreased cortical activation and disengagement from sustained processing of positive information. This pattern aligns with the anhedonic characteristics of MDD patients, where attentional resources are not effectively maintained toward rewarding stimuli~\cite{pizzagalli2014depression}.
\subsubsection{Correlation Between Model Logits and Neural Features}
\begin{figure}[t!]
 \includegraphics[width=\columnwidth]{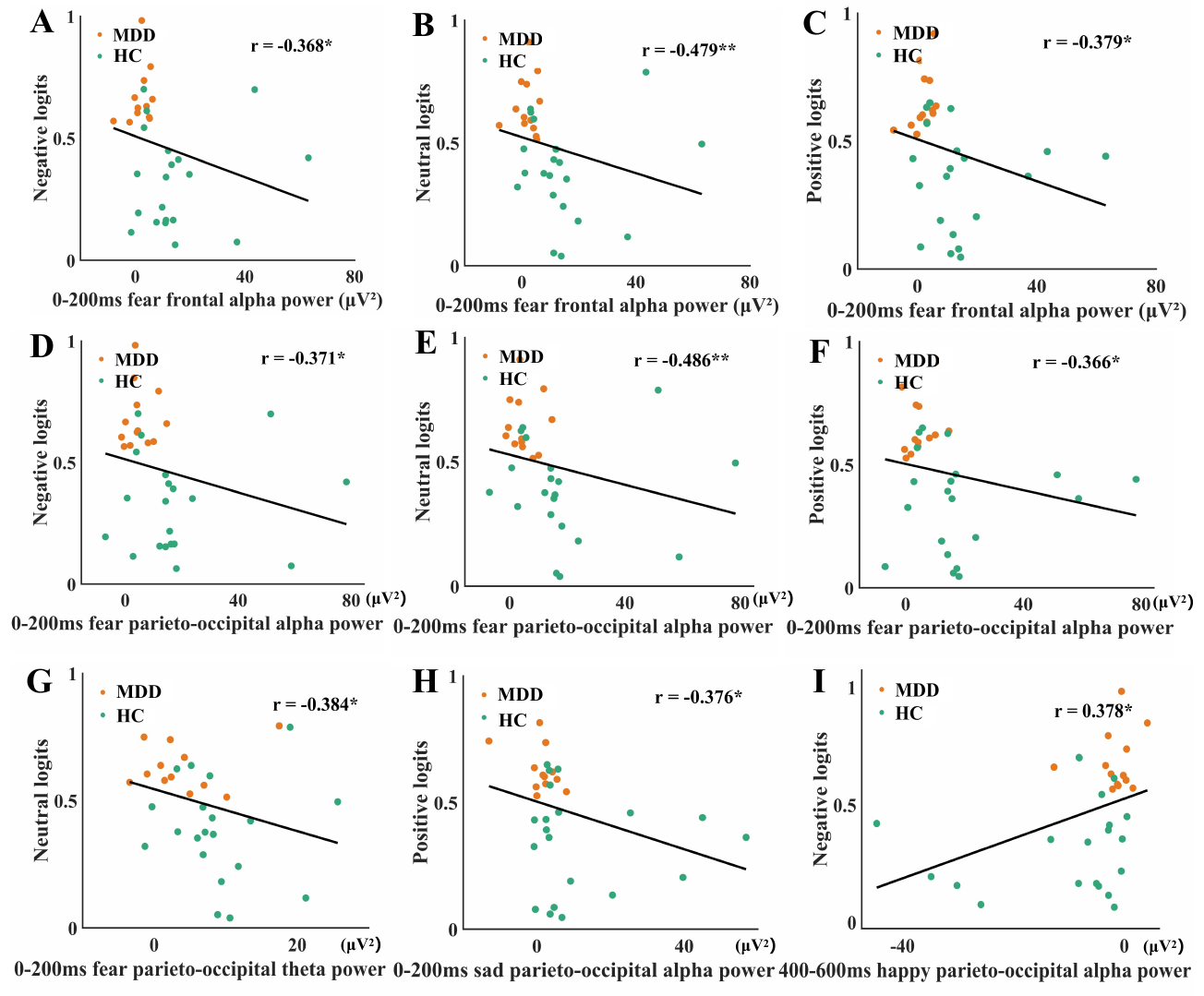}
 \caption{
 Spearman correlations between model-derived depression logits and time-frequency power. (A-C) In the frontal region (0-200 ms, fearful condition), alpha power was negatively correlated with logits from the negative, neutral, and positive speech conditions. (D-F) In the parieto-occipital region (0-200 ms, fearful face condition), alpha power showed significant negative associations with logits under the negative, neutral, and positive speech conditions. (G) Parieto-occipital theta power (0-200 ms, fearful condition) negatively correlated with neutral speech logits. (H) Under the sad face condition (0-200 ms), parieto-occipital alpha power negatively correlated with positive speech logits. (I) In the happy face condition (400-600 ms), parieto-occipital alpha power positively correlated with logits from the negative speech condition. Each dot represents one participant (orange: MDD; green: HC).
 % Spearman correlations between model-derived depression logits and time-frequency power.
 }\label{fig:logits}
\end{figure}
%
% To assess whether the speech-based diagnostic model’s depression estimates align with neural indices of depressive symptomatology, we examined the correlation between model logits and theta/alpha power shown significant differences between two groups. Model-derived estimates represent the predicted probability of being classified as depressed, ranging from 0 to 1.
To assess whether speech-based diagnostic model estimates align with neural indices of depressive symptomatology, we examined correlations between model logits (predicted depression probability, ranged 0-1) and theta/alpha power that showed significant group differences (Fig.~\ref{fig:logits}).

We found that frontal alpha power during fearful face processing was negatively correlated with depression logits across all speech conditions (negative: $r = -0.368$, $p = 0.036$; neutral: $r = -0.479$, $p = 0.005$; positive: $r = -0.379$, $p = 0.030$; Fig. \ref{fig:logits}A-C). Similarly, parieto-occipital alpha power showed negative correlations with logits from all speech conditions during fearful faces (negative: $r = -0.371$, $p = 0.034$; neutral: $r = -0.486$, $p = 0.005$; positive: $r = -0.366$, $p = 0.037$; Fig.~\ref{fig:logits}D-F). These findings indicate that reduced frontal and parieto-occipital alpha activity during early processing of fearful faces reliably co-varies with higher model-predicted depression likelihood.
Additionally, parieto-occipital theta power during fearful face processing was negatively correlated with neutral speech logits during fearful faces ($r = -0.384$, $p = 0.028$; Fig.~\ref{fig:logits}G).
Under sad face conditions, parieto-occipital alpha power in the 0-200 ms window was negatively correlated with positive speech logits ($r = -0.376, p = 0.032$; Fig.~\ref{fig:logits}H).
These findings indicate that speech model-predicted depression probability is consistent with the significant reduction in early neural activity during negative face processing.
During happy face processing, parieto-occipital alpha power in the 400-600 ms window positively correlated with negative speech logits ($r = 0.378$, $p = 0.031$; Fig.~\ref{fig:logits}I).
This result indicates that higher speech model-predicted depression probability is associated with increased later neural activity during positive face processing.
\subsubsection{Correlation Between $\Delta$Logits and EEG}
\begin{figure}[t!]
 \centering
 \includegraphics[width=\columnwidth]{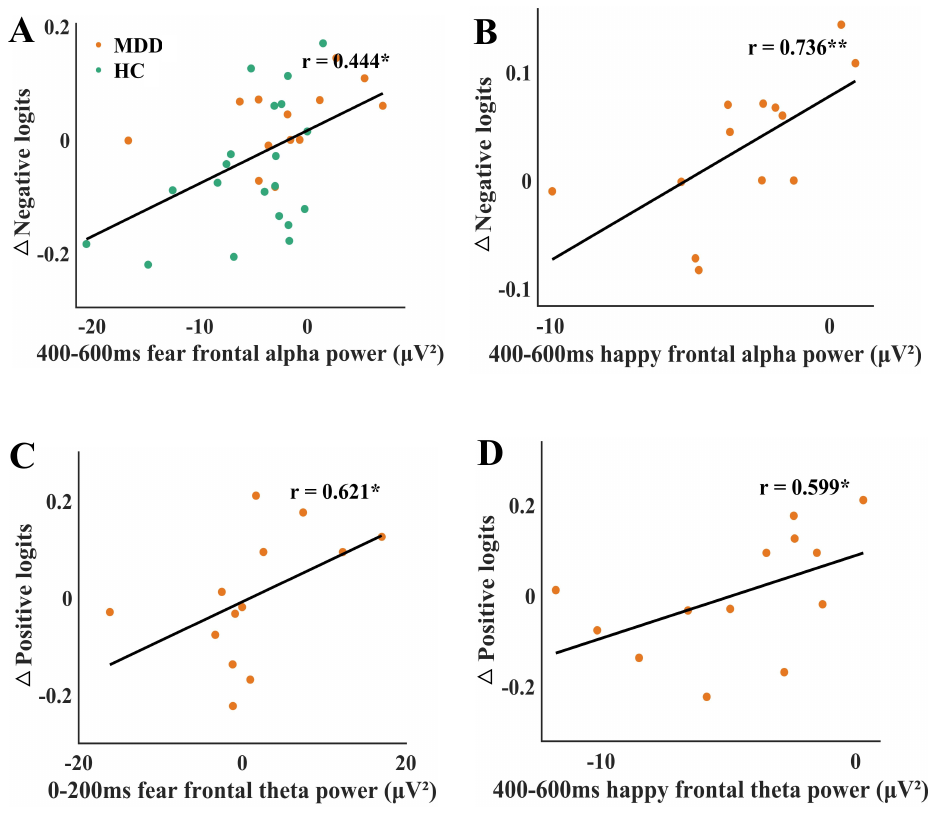}
 \caption{
 Spearman correlations between $\Delta$logits (positive minus neutral and negative minus neutral) and time-frequency power. (A) Frontal alpha (400-600 ms, fearful condition) positively correlated with negative $\Delta$logits across groups ($r = 0.444$, $p = 0.010$). (B) In MDD, frontal theta (0-200 ms, fearful condition) positively correlated with positive $\Delta$logits ($r = 0.621$, $p = 0.027$). (C) In MDD, frontal alpha (400-600 ms, happy condition) positively correlated with negative $\Delta$logits ($r = 0.736$, $p = 0.006$). (D) In MDD, frontal theta (400-600 ms, happy condition) positively correlated with positive $\Delta$logits ($r = 0.599$, $p = 0.034$). Each dot represents one participant (orange: MDD; green: HC).
 % Spearman correlations between $\Delta$logits ($\Delta$positive and $\Delta$negative) and time-frequency power.
 }\label{fig:delat_logits}
\end{figure}
% While the previous analysis focused on absolute depression scores under different speech conditions, we further explored how these estimates shift relative to the neutral baseline to capture potential context-specific neural modulation. 
% To this end, we computed $\Delta$logits by subtracting the neutral speech logits from those of the positive and negative speech conditions. 
% Spearman correlation analyses were then performed between these $\Delta$logits and EEG power features across time-frequency representations of different time windows (early/late), frequency bands (alpha/theta) and regions (frontal/parieto-occipital).
To further investigate the relationship between emotional expression in speech and potential context-specific neural modulation, we calculated Spearman correlation between $\Delta$logits ($\Delta$positive and $\Delta$negative) and EEG power features across time-frequency representations of different time windows (early/late), frequency bands (alpha/theta) and regions (frontal/parieto-occipital). 
Significant correlations were found between $\Delta$logits and mean time-frequency power (Fig. \ref{fig:delat_logits}). When collapsing across groups, $\Delta$negative were positively correlated with frontal alpha power between 400 and 600 ms during the fearful condition ($r = 0.444$, $p = 0.010$; Fig. \ref{fig:delat_logits}A). This indicates that, regardless of MDD patients or HCs, individuals who tend toward more negative linguistic expressions invest more cognitive resources in late-stage processing of fearful stimuli.

However, group-specific analyses revealed unique correlation patterns in MDD patients that were absent in HCs. Within the MDD group, $\Delta$negative showed a strong positive association with frontal alpha power during the happy condition in the later time window from 400 to 600 ms ($r = 0.736$, $p = 0.006$; Fig. \ref{fig:delat_logits}B). This finding reveals that in MDD patients, stronger negative linguistic tendencies are associated with enhanced frontal alpha activity when processing positive stimuli. 
Additionally, $\Delta$positive was significantly correlated with frontal theta power during fear (0-200 ms: $r = 0.621$, $p = 0.027$; Fig. \ref{fig:delat_logits}C) and happy conditions (400-600 ms: $r = 0.599$, $p = 0.034$; Fig. \ref{fig:delat_logits}D). These findings suggest that in MDD patients, more positive speech expressions are associated with enhanced frontal theta activity, reflecting the increased cognitive effort needed for top-down control mechanisms when producing positive emotional expressions.
% This pattern aligns with our hypothesis that emotional arousal, rather than valence polarity, serves as the critical factor in emotional expression within speech.
%
%

% \section{Discussion}

\section{Discussion}~\label{sec:dis}
While the general state of depression is often characterised by low valence and low arousal according to the circumplex model of affect~\cite{posner2005circumplex}, the high-arousal tasks used in this study (eliciting positive and negative affect) measure the acute response to an emotional provocation~\cite{girard2014nonverbal} rather than this general state. A central finding of this study is the strong empirical support for the emotional arousal hypothesis in speech. This acute response effectively exposes the underlying emotional dysregulation in depression, such as positive attenuation~\cite{disner2011neural} and the amplification of negative affect~\cite{gotlib2004attentional,gotlib2010cognition}. In contrast, the low-arousal neutral task provides insufficient emotional provocation, leading to less distinct speech patterns between the groups. Our findings confirm that the key acoustic differentiator was not the polarity of the emotion (valence), but the atypical manner in which individuals with depression responded to any high-arousal prompt.

The successful cross-linguistic generalisation of the CDMA framework warrants a careful distinction from transfer learning approaches. Transfer learning demonstrates the data-driven engineering feasibility of knowledge transfer by forcing a statistical alignment between source and target domains, a process often reliant on a large source dataset~\cite{pan2009survey}; thus, its success in depression detection~\cite{qin2025cross} may depend heavily on the transferred knowledge. In contrast, the success of the CDMA framework stems not from inheriting knowledge but from its inherent architectural design. This provides a crucial inductive bias~\cite{battaglia2018relational} that allows it to naturally discover the shared pattern of depression even within a typologically distinct language. This study therefore not only presents a framework effective on smaller datasets where large-scale source data is unavailable but also scientifically validates that there are generalizable principles to be learned in the first place, which provides a theoretical foundation for transfer learning~\cite{yosinski2014transferable}. This suggests that applying transfer learning to the CDMA framework in the future would likely be highly successful, transforming the process from a blind alignment into a guided fine-tuning of already meaningful features.

Our neurophysiological analysis provides the direct answer to the question of whether capturing acoustic signals that have a genuine neurophysiological basis, or by merely exploiting superficial statistical artifacts unrelated to the underlying pathophysiology of depression. 
EEG results revealed distinct group differences between patients with MDD and HCs in both the theta and alpha frequency bands during the early encoding phase of negative face, highlighting alterations in the neural dynamics of emotional processing in depression. Consistent with the attentional bias to negative information previously reported in MDD~\cite{disner2011neural}, these attenuated oscillatory activities suggest abnormally attentional engagement toward negative stimuli, further indicating deficient top-down cognitive control~\cite{heller2009reduced} and impaired regulation of emotional responses at early sensory stages~\cite{paquet2022sensory}. In contrast, positive emotional processing showed no early group differences, but patients with MDD displayed increased parieto-occipital alpha power during later processing. This enhancement of alpha activity implies decreased cortical activation and withdrawal of attention from happy faces~\cite{klimesch2007eeg}, aligning with reduced motivation toward positive stimuli commonly observed in depression~\cite{epstein2006lack}.
Importantly, correlations between EEG oscillations and speech-based model logits provide crucial physiological validation for the computational model. The negative correlations between model-predicted depression likelihood and early frontal and parieto-occipital alpha power during negative-face processing indicate that individuals classified as more depressed by the model also showed weaker early neural responses to emotional stimuli. And the positive correlation between model logits and later alpha power during happy-face processing suggests that higher predicted depression levels are linked to greater neural inhibition during positive emotion processing. Furthermore, regardless of the valence, stronger tendencies of predicted depression were associated with increased frontal alpha and theta activity during emotion processing in patients with MDD, suggesting neural evidence for the critical role of emotional arousal in speech-derived identification of depression.
Therefore, the cross-linguistic robustness might be partly explained by the observed neurophysiological validation, however, it is crucial to acknowledge that without corresponding EEG data from the another corpus, this link remains speculative and requires further multimodal cross-lingual investigation.

Despite these contributions, several limitations should be acknowledged, which also point toward avenues for future research. First, our EEG analysis utilised the MODMA dataset. While this is the largest publicly available dataset of its kind, the number of participants with high-quality concurrent recordings remains relatively modest. Replicating these correlational findings in a larger, independent cohort is therefore an essential next step. Second, another limitation is the correlational nature of our findings. We cannot infer causality from our data due to potential confounding variables such as medication status and specific comorbidities (e.g., anxiety disorders), which were not reported in the original dataset publication. Although the original study applied rigorous diagnostic and exclusion criteria, the absence of this specific information means that their potential influence on speech and EEG features could not be assessed. For instance, a psychomotor slowing might simultaneously lead to reduced speech variability and altered alpha/theta band power, creating the observed correlation without a direct causal link between the two. Future research with longitudinal designs would be necessary to probe potential causal pathways.
\section{Conclusion}\label{sec:con}
This study makes three primary contributions to computational mental health assessment. 
First, we provide the first empirical validation that the emotional arousal, rather than valence polarity, is the critical factor enhancing detection performance, with both positive and negative spontaneous speech significantly outperforming neutral speech.
Second, we provide strong evidence for the cross-linguistic generalizability of the CDMA framework for speech-based depression detection, achieving comparable performance on Chinese Mandarin speech (F1-score up to 89.6\%) as previous Italian validation. 
Third, we establish the first neurophysiological validation of a speech-based depression model by demonstrating significant correlations between model-derived depression estimates and established neural markers of emotional dysregulation.

The clinical implications of this research are substantial. The high recall performance (95.7\%) minimises false negatives, crucial for depression screening, while the cross-linguistic robustness enables potential deployment in multilingual clinical environments. Future work could also investigate the integration of other physiological signals, such as heart rate and skin conductance, to further enrich the model's understanding of a patient's mental state and move closer to a holistic, computationally-driven assessment of depression.

\bibliographystyle{IEEEtran}
\bibliography{submission}
\end{document}